\documentclass[aps,prd,preprint,groupedaddress,superscriptaddress,nofootinbib,longbibliography,showpacs,showkeys,bibnotes]{revtex4-1}
\usepackage{amsmath}   
\usepackage{amsfonts}
\usepackage[latin1]{inputenc}
\usepackage{graphicx}  
\usepackage[usenames,dvipsnames]{xcolor}
\usepackage{hyperref} 
\hypersetup{colorlinks=true, citecolor=Blue, linkcolor=Blue, urlcolor=Blue}

\begin{document}

\title{Interpretations and naturalness in the radiation-reaction problem}

\author{Carlos Barcel\'o}
\email[]{carlos@iaa.es}
\affiliation{Instituto de Astrof\'isica de Andaluc\'ia (IAA-CSIC), Glorieta de la Astronom\'ia, 18008 Granada, Spain}

\author{Luis J. Garay}
\email[]{luisj.garay@ucm.es}
\affiliation{Departamento de F\'{\i}sica Te\'orica and IPARCOS, Universidad Complutense 
de Madrid, 28040 Madrid, Spain}
\affiliation{Instituto de Estructura de la Materia (IEM-CSIC), Serrano 121, 28006 Madrid, Spain}

\author{Jaime Redondo-Yuste}
\email[]{jairedon@ucm.es}
\affiliation{Departamento de F\'{\i}sica Te\'orica and IPARCOS, Universidad Complutense 
de Madrid, 28040 Madrid, Spain}


\begin{abstract}
After more than a century of history, the radiation-reaction problem in classical electrodynamics still surprises and puzzles new generations of researchers. Here we revise and explain some of the paradoxical issues that one faces when approaching the problem, mostly associated with regimes of uniform proper acceleration. The answers we provide can be found in the literature and are the synthesis of a large body of research. We just present them in a personal way that may help in their understanding. Besides, after the presentation of the standard answers we motivate and present a twist to those ideas. The physics of emission of radiation by extended charges (charges with internal structure) might proceed in a surprising oscillating fashion. This hypothetical process could open up new research paths and a new take on the equivalence principle.

\end{abstract}

\maketitle
\tableofcontents

\section{Introduction}

At the end of 19th Century, physicists realised that accelerating charges should emit electromagnetic radiation and, as a consequence, there should be some back-reaction acting upon them (see e.g.~\cite{McDonald1998} for how this notion came to the physics forefront). Since then, the so-called classical electromagnetic radiation-reaction problem has been renovating once and again as an attractive problem full of controversies and insights touching central topics in physics. And all that without invoking competing paradigms (in Kuhn's terminology), just using the standard Maxwell field equations. As of today, it is fair to say that there are still several aspects of the problem which do not have a completely satisfactory understanding.

In addition to the intrinsic interest of the classical electromagnetic radiation-reaction problem, in modern times a renovated interest on it comes about from two closely related phenomena: the Unruh effect ---interaction between accelerated quantum detectors and quantum fields--- and the gravitational radiation-reaction problem. On the one hand, the physics of accelerated quantum detectors leads to some controversial interpretational questions analogous to those with accelerated charges, e.g. surprisingly, in a first look a uniformly accelerated Unruh-de Witt detector does not produce any radiation~\cite{Grove1986,Massaretal1993,Crispinoetal2008}. Whether this is the case or not can have important consequences in understanding for example the Hawking emission by black holes~\cite{Barbadoetal2016}. 
On the other hand, the trajectory of a small star or black hole, with $m \neq 0$, attracted by a supermassive black hole, with $M \gg m$, differs from the geodesic it would have followed in the test-mass limit owing to the emission of gravitational waves (see e.g.~\cite{BarackPound2019,Olteanetal2020} and references therein). The calculation of the back-reacted trajectories has become an important problem in gravitational wave astronomy, since these types of situations are expected to be observable sources of gravitational waves~(for a review on Extreme Mass Ratio Inspirals see e.g.~\cite{Amaro2018}.) To better understand these arguably more complicated problems it is sensible to take one step back and clearly understand the classical electromagnetic problem.

Our humble intention with the present work is to help clarifying a selection of questions one can naturally ask oneself when thinking about the classical electrodynamic radiation-reaction problem. Answers to most of these questions are already present in the relevant literature but sometimes not explicitly or clearly enough to stop being a source of confusion. In addition, we will show that some of these answers are not as compelling as they may seem, leaving still holes for further exploration.

In this paper we will always have in mind a charged object as an structured extended entity which is however very small from the point of view of the observational parameters on the laboratory. For example, we can think of a macroscopic grain of dust with a net charge. For many characteristics of its behaviour, but not all, it can be treated as a point-like object. Whether the findings we shall discuss apply in some way to elementary particles such as the electron is more difficult to know. On the one hand, in many respects their behaviour is deeply quantum. On the other hand, at the current experimental level they do not show any structure. In any case, we consider the classical understanding as a rich conceptual toolkit.

Let us start by writing down an itemised number of questions that surely many readers have come about when thinking about the radiation-reaction problem. Then, each section will be devoted to clarify each of them (relevant references will be given in the corresponding sections).
\begin{itemize}

\item
Does a charge restrained from falling in a gravitational well, so that it remains static, radiate? The conceptual problem arises because of an interpretational clash. On the one hand, people are typically convinced that an accelerated charge in Minkowski spacetime emits radiation towards the asymptotic regions. On the other hand, people are typically inclined to believe that a charge at rest in their desktop is not radiating towards infinity and so it does not require a continuous supply of energy. But a charge at rest in a gravitational field is locally accelerating so a tentative application of the equivalence principle suggests that it should emit\ldots 

\end{itemize}

Before putting forward the next questions we need to recall the structure of the Lorentz-Abraham-Dirac (LAD) self-force~\cite{Lorentz1892,Abraham1902,Dirac1938,Rohrlich1997}. 
The well-known LAD expression for the self-force acting on a point-like particle has the form 
\begin{align}
F_{\rm S}^b=-m_{\rm ed} a^b + \frac23 q^2\left( \frac{d a^b}{d \tau}-(a^c a_c) u^b \right),
\label{R-LAD}
\end{align}
which becomes 
\begin{align}
{\mathbf F}_{\rm S}= -m_{\rm ed}{\mathbf a} + \frac23 q^2 \dot{\mathbf a}
\label{NR-LAD}
\end{align}
in the non-relativistic limit. In the relativistic expression $u^b$, $a^b$ represent the four-velocity and four-acceleration, respectively; $\tau$ is the proper time of the trajectory; $q$ is the charge of the particle; and $m_{\rm ed}$ is an electrodynamic mass whose value encodes the specific electrodynamic energy carried by the charge. Boldface symbols are used to represent spatial vectors in the non-relativistic equation, whose components will be labelled by Latin indices $i,j,k\ldots$ when necessary.   

The first term in \eqref{R-LAD} is typically absorbed in a renormalised mass for the point-like system that then contains some electrodynamic contribution. In this way the actual dressed mass of the charge consists of a bare mass plus an electrodynamic contribution: $m_{\rm D}=m_{\rm B}+m_{\rm ed}$. 
In the point-like limit this electrodynamic mass would be divergent but for a real extended system it would be finite and dependant on the internal structure of the system. Thus, very frequently one forgets about this term leaving as the actual self-force just
\begin{align}
\tilde{F}_{\rm S}^b=\frac23 q^2 \left( \frac{d a^b}{d \tau}-(a^c a_c) u^b \right),
\label{Ren-LAD}
\end{align}
whose nonrelativistic version is
\begin{align} 
\tilde{\mathbf F}_{\rm S}=   \frac23 q^2 \dot{\mathbf a}.
\label{Ren-LAD-nonrel}
\end{align}
As we will see below this might lead to some interpretational difficulties.

\begin{figure}
	\centering
	\includegraphics[width=.3\textwidth]{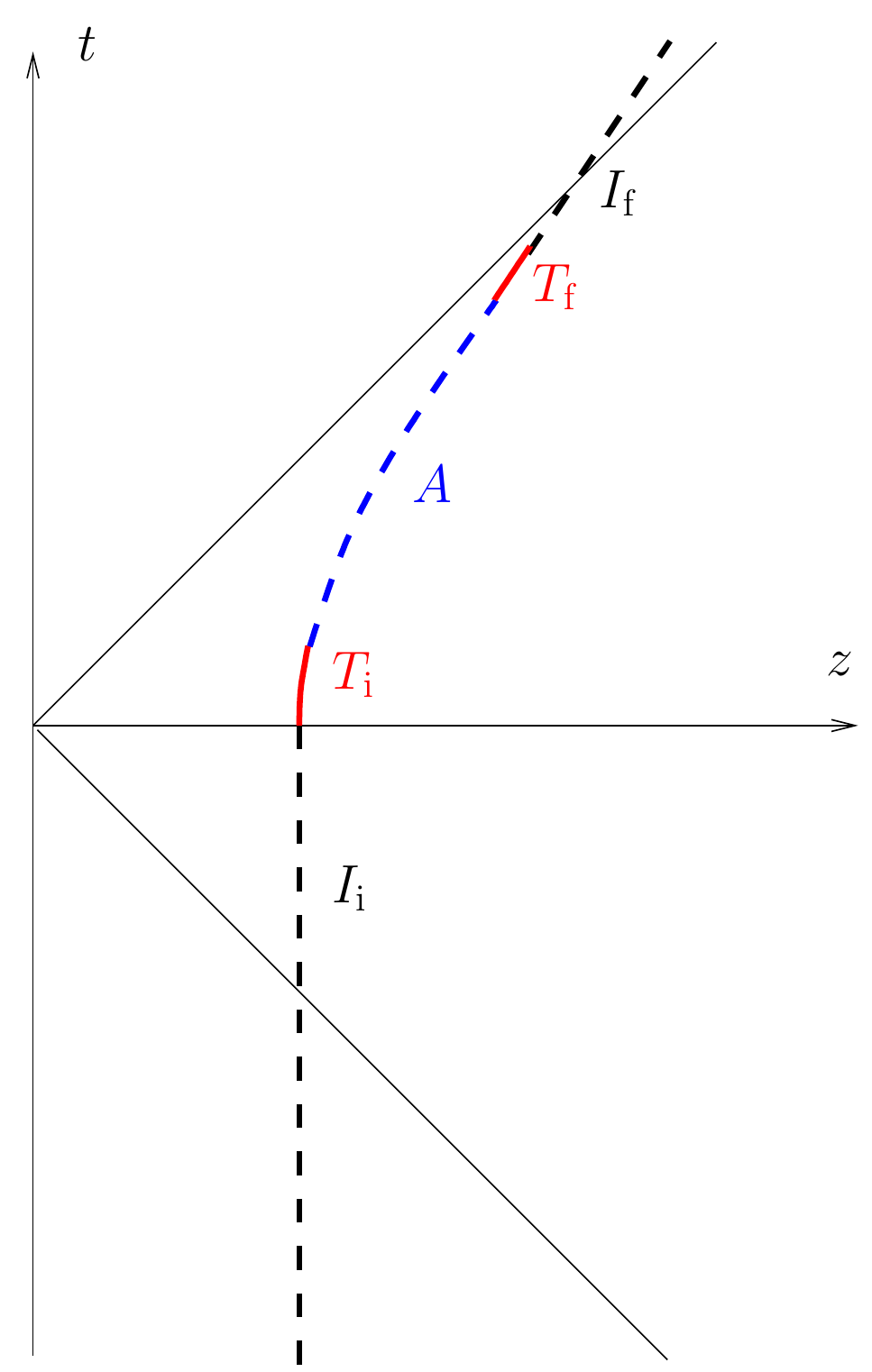}
	\caption{Accelerated trajectory of a particle in Minkowski spacetime. The particle starts from rest (black dashed line); then it passes through a relatively short period of non-uniform acceleration (red thick line); then through a uniformly accelerating regime (blue dashed line); then again through a brief transient regime (red thick line); finally the acceleration disappears and the particle remains with inertial motion (black dashed line).}
	\label{Fig:Regimes}
\end{figure}

It is well known that the LAD force leads to unphysical solutions (i.e. pre-accelerating and run-away solutions) given the third-order nature of the resulting dynamical equation~\cite{FultonRohrlich1960}. But it is also well known that this equation is just an approximation to a more appropriate second order equation devoid of these unphysical solutions~\cite{Rohrlich2007,Yaghjian1992,Rohrlich1997}. However, with a bit of care people can and actually do continue using the LAD self-force to interpret radiation-reaction phenomena. For instance, it appropriately takes care of the energy budget in standard physical situations. We pay a prize though: with expression~\eqref{Ren-LAD} we face at least three interpretational problems, listed in the following.
\begin{itemize}
\item
Using the LAD force, the total amount of work done in a process in which the charged particle starts and ends in inertial motion is precisely equal to the growth of kinetic energy plus the total amount of radiated energy. The self-force part alone~\eqref{Ren-LAD} is responsible for the radiated energy.  One can check, starting from $d E_{\rm S}/d\tau=F_{\rm S}^0=F_{\rm S}^i u_i/u^0$, that the work done by the self-force is given by
\begin{align}
\Delta E_{\rm S}= \frac23 q^2 \int  \left( \frac{d a^i}{d \tau}-(a^c a_c) u^i \right) u_i \gamma^{-1}d\tau,
\label{Budget1}
\end{align}
where $\gamma=u^0$ is the Lorentz factor. By performing some straightforward manipulations, the integral above can be rearranged in two terms, so that
\begin{align}
\Delta E_{\rm S}= \frac23 q^2 \int  \frac{d}{d \tau}(a^i u_i\gamma^{-1}) d\tau
-\frac23 q^2 \int (a^c a_c) dt.
\label{Budget2}
\end{align}
The first term is the integral of a total derivative and therefore vanishes for trajectories that start and end with zero acceleration. The second term is precisely the total energy lost by the system by radiation emission (Larmor's relativistic formula~\cite{Larmor1897,Heaviside1902}, and see below).  
In this way we see that the energy budget appears to be correctly taken care of. 

Now, consider a situation separated into five different and consecutive regimes (see figure~\ref{Fig:Regimes}): one initial inertial regime that we will denote $I_{\rm i}$; a transient in which some acceleration is established, $T_{\rm i}$; one arbitrarily long period of uniform acceleration, $A$; another transient in which the acceleration disappears, $T_{\rm f}$; and another final inertial regime, $I_{\rm f}$. The LAD expression suggests that all the work done by the self-force on the system takes place during the transients $T_{\rm i}, T_{\rm f}$, even though most of the radiation has been emitted during stage $A$. This situation is interpretationally difficult and can make one think that there might be local violations of the energy budget which however do not lead to any global failure (this puzzle is described e.g. in~\cite{Grallaetal2009}).

\item
While the radiation emission in a regime of uniform acceleration is stationary, the self-force vanishes. It seems that the emission of radiation in that regime is not influencing in any way the trajectory of the charge which seems to be driven only by the external force.
It might appear that it does not take any more effort to move a particle when it is charged that when it is not (of course for equal masses). 

\item
The LAD self-force has another interpretational problem. When an acceleration is established, as in transient $T_{\rm i}$, the back-reaction appears to go in favour of this very acceleration, the opposite to what one might have expected.  

\end{itemize}

In the following sections we shall answer all these questions. In section~\ref{Sec:Radiation}, we will discuss the issues associated with the emission of radiation. Then, section~\ref{Sec:Selfforce} will deal with the problems associated with the self-force. As already mentioned, the answers found in these sections can be found in the literature. Our contribution here is to collect them to construct a compelling interpretation of all the issues at stake. Later, in section~\ref{Sec:Twist} we introduce a twist in the previous discussion, suggesting a potential change on how radiation-reaction proceeds. In the final section we provide a short summary of the paper and some concluding remarks.

\section{Radiation by uniformly accelerating charges}
\label{Sec:Radiation}

In the late nineteenth century it was already asserted that an accelerated charge should emit electromagnetic radiation~\cite{Larmor1897,Lorentz1892}. However, this apparently clear idea was subjected to intense debate for many years (some central references are~\cite{Born1909,Pauli1958,vonLaue1919,Schott1912,Schott1915,Milner1921,Bondietal1955,Feynman1962,FultonRohrlich1960,Boulware1980}). In this section we comment on some of the core questions on this debate. 

\subsection{Does a uniformly accelerating charge in Minkowski spacetime radiate?}

Let us start by mentioning that in order to analyse this question one can deal with idealised point particles.
Indeed, on the one hand, the linearity of Maxwell equations allows to deal with distributional sources. On the other hand, the radiation field shows up at large distances from the source; so the divergences of the field at the point particle position should not cause any trouble when analysing its radiative characteristics.

In favour of the assertion that an accelerating charge radiates, there is the direct argument of calculating the fields generated by a moving point charge based on retarded Li\'enard-Wiechert potentials~\cite{Lienard1898,Wiechert1901,Jackson2012}.
The Pointing-vector flux through a sphere at infinity can be calculated resulting in the radiation rate
\begin{equation}
    \mathcal{R}=\frac{2}{3} q^2 a^b a_b.
\end{equation}
This is Larmor's relativistic formula with $a^b$ the standard four-acceleration that measures any deviation from inertial motion. For a given trajectory with proper constant acceleration $g$, a hyperbolic motion in Minkowski spacetime, $a^b a_b=g^2=$ constant. Therefore, a straightforward interpretation of the previous formula is that for a uniform acceleration one would have a constant emission rate.

Arguments against this interpretation were put forward since the very beginning by researchers such as Born~\cite{Born1909} and Pauli~\cite{Pauli1958}. Many other such as von Laue \cite{vonLaue1919}, and later Hill~\cite{Hill1947} and Feynman~\cite{Feynman1962}, subscribed and elaborated on these arguments. Essentially, on the one hand there is Pauli's argument. It is based on the fact that on the hypersurface $t=0$, where the hyperbolic trajectory passes through its point of zero velocity, the magnetic field vanishes. So it seems impossible to associate a wave zone and a non-vanishing Pointing vector to the process. The problem with this argument appears to have been first identified by Drukey \cite{Drukey1949} and then further cleared up by Bondi, Gold and Spencer~\cite{Bondietal1955} and Fulton and Rohrlich~\cite{FultonRohrlich1960}. The problem is that, in order to identify the radiation produced at a point of the trajectory, one has to analyse the limit of large spheres $R \to \infty$ within the causal lightcone $R=t-t_{\rm emission}$. Only with the values of the magnetic field in one spacelike hypersurface one cannot know whether there is radiation or not at infinity. In geometrical language, one has to analyse the structure of null infinity and not of spatial infinity. On the other hand, Born's argument is based on the conformal invariance of Maxwell's equations: if there is no radiation when a particle is at rest, there cannot be when the particle is uniformly accelerating, as this movement can be attained by a special conformal transformation. The problem with Born's argument is that he was using, without realising it, not just the retarded fields of a single charge but a combination of half-advanced plus half-retarded fields associated with two mirror charges. This field combination indeed does not lead to radiation at infinity. However, this field solution does not represent the physical situation one is interested in. In fact, this field solution is the result of applying a special conformal transformation to the Coulomb field of a particle at  rest~\cite{Milner1921,FultonRohrlich1960,Fultonetal1962}. Against the conformal invariance argument, we could say that the particular solution, one single accelerated charge with just retarded potentials, spontaneously breaks conformal invariance. 

We guess that the idea that uniformly accelerated charges could not radiate was favourably taken by many people in part because they found that it was consistent with the fact that in these trajectories the LAD self-force \eqref{Ren-LAD} vanishes (we will start discussing this problem at the end of this section and continue in the next). 
For instance, in a sufficiently small neighbourhood around a uniformly accelerating charge (a world tube surrounding the charge trajectory) one realises that the retarded electromagnetic fields do not exhibit any specific retarded characteristics~\cite{Boulware1980}: locally the retarded field is equal to the advanced field. In fact, this observation alone could be used to predict that the self-force should vanish for uniform acceleration.
It is not as if something physical is being emitted locally by the charge (as one would imagine the emission of a photon). The radiative characteristics are appreciated only far from the source and, as we will see, take into account global properties of the spacetime. 

The answer to the question in the title of this subsection is yes, a charge subject to uniform acceleration in Minkowski spacetime radiates, but this assertion should always go hand by hand with further qualifications, as we are about to explain.

\subsection{Does a charge restrained from falling towards a gravitational potential well radiate?}

The idea that when a particle accelerates in Minkowski spacetime it radiates is relatively easy to swallow. Then, it might appear that by looking at whether a particle radiates or not one could distinguish whether its behaviour is inertial or not. 

What happens when a charge is kept from falling towards a potential well remaining static (either by some rocket or by being on top of a solid surface attached to a planetary structure)? The principle of equivalence seems to tell us that this situation should be indistinguishable from an acceleration in Minkowski spacetime (at least locally, without considering inhomogeneities in the gravitational field). 
However, it is difficult to imagine that an observer at rest with respect to the charge will observe radiation as for him the structure of the fields surrounding the charge is static. The same applies to any other observer at rest with respect to the generator of the  gravitational field, including those at infinity. Therefore, there should be no radiation escaping to infinity. If this is the case, there seems to be a problem with the equivalence principle: by measuring whether a charge radiates or not one could distinguish whether it is accelerating in Minkowski spacetime or experiencing a uniform gravitational field. 

The solution to this puzzle was provided by Boulware~\cite{Boulware1980}, elaborating on previous works by Fulton and Rohrlich~\cite{FultonRohrlich1960} and Coleman~\cite{Coleman1961}. Regarding the equivalence principle, the situation that should be compared with the charge at rest in the gravitational field is that of an observer following the accelerating charge in Minkowski spacetime (a comoving accelerating observer). The presence of Rindler horizons in this case makes this observer unable to feel any radiation. Boulware's argument is that by looking at the fields on the right wedge of Rindler spacetime, one cannot distinguish between retarded and advanced solutions. For instance, one could perfectly think that the solution contained half retarded plus half advanced fields (as in Born's argument), which would entail no radiation at infinity, and hence no overall self-force. As a final conclusion this work advances the thesis that the presence of radiation is observer dependent. 

To explicitly check that a charge restrained from falling in a gravitational well does not radiate, let us provide here a simple calculation based on Rindler spacetime. We are using the following  system to completely separate the problem at hand from issues related with the presence of tails in the propagators in curved spacetimes. Rindler spacetime can be interpreted as representing the uniform gravitational field that observers would perceive when moving in a small region close to the surface of a very large star or black hole~\cite{Rindler1960}. Take Schwarzschild metric in Schwarzschild coordinates, write $r= 2M+h$, and make the approximation $h\ll 2M$. The approximate metric reads
\begin{equation}
ds^2 = -\frac{h}{2M} dt^2 + \frac{2M}{h}dh^2 + (2M)^2 d\Omega_2^2. 
\end{equation}
Using the coordinate $z=2\sqrt{2M~h}$ and local transverse Cartesian coordinates $x,y$, we can write this metric as
\begin{equation}
ds^2= -g^2 z^2 dt^2 +dz^2 + dx^2 + dy^2, 
\end{equation}
with $g=1/(4M)$ being the surface gravity of the black hole. One can think of this metric as the right wedge of Minkowski spacetime written in Rindler coordinates. However, here we go one step further and consider as our global metric a spacetime consisting of two Rindlerian wedges of Minkowski spacetime pasted together through a thin membrane. This amounts to consider two copies of the previous metric pasted at $z=0$. It is not difficult to check that this global metric is now a solution of Einstein equations with a diagonal stress-energy tensor (SET) of the form
\begin{equation}
\{\rho,p_z,p_x,p_y\}=\{ 0, 0, -2g,-2g\}\delta(z).
\end{equation}
Therefore, it is not empty and is globally different from Minkowski spacetime (two Rindler wedges have been cut out from it). In fact, this geometry can be understood as a limiting situation within the family of symmetric Schwarzschild thin-shell wormholes~\cite{Visser1995}. One Schwarzschild thin-shell wormhole can be sustained by a thin shell located at radius $a$ and having surface density and transverse tensions
\begin{equation}
\sigma=-\frac{1}{2\pi a} \sqrt{1-\frac{2M}{a}},\qquad
\theta=-\frac{1}{4\pi a} \frac{1-\frac{M}{a}}{\sqrt{1-\frac{2M}{a}}}.
\end{equation}
If we take the neck to be located at $a=2M+\epsilon$, $\epsilon \ll 2M$, and take the limit $M \to +\infty$ at the same time that $\epsilon \to 0$ keeping $\epsilon M$ constant and finite, then one obtains precisely the previous Rindlerian geometry with $(\epsilon M)^{-1}=2 (8\pi)^2 g^2$. The density term gets diluted to zero in the limiting  process; not so the tension terms. In the following, when making a calculation in Rindler spacetime we will have in mind this spacetime.

Let us consider a static charge fixed at a distance $z_0$ from the domain wall above. The calculation of the four-potential in the Lorenz gauge yields 
\begin{align}\label{potential}
    A_t&=-\frac{qg^2}{4\pi} \frac{\rho^2}{\xi}, \qquad
    A_z=-\frac{q}{4\pi(z-z_0)}, \qquad A_x=A_y=0,
\end{align}
where $\rho^2=x^2+y^2+(z-z_0)^2+g^{-2}$ and $\xi^2=g^2\rho^4-4(z-z_0)^2$. The electromagnetic field  $F_{ab}=\partial_a A_b-\partial_b A_a$  is easily computed to find:
\begin{align} 
        F_{t x}&=-\frac{2q g^2}{ \pi} \frac{ x(z-z_0)^2}{\xi^{3}}, \qquad
        F_{t y}=-\frac{2q g^2}{ \pi } \frac{ y(z-z_0)^2}{\xi^{3}},\nonumber\\
        F_{t z}&=\frac{q g^2 }{ \pi } \frac{ (z-z_0)[\rho^2-2(z-z_0)^2]}{\xi^{3}} ,\nonumber\\
        F_{xy}&=F_{xz}=F_{yz}=0 .
\end{align} 

At large distances from the domain wall one finds a behaviour 
\begin{equation}
    F_{t z}\to-\frac{qg}{2\pi(z-z_0)}.
\end{equation}
This could be taken as an indication of the presence of radiation. However, the magnetic field is exactly zero, and therefore, so is the Poynting vector measured by an observer far away from the source. For this observer the four-velocity is $u^a=(1,0,0,0)$ and the Poynting vector $S^i=T^{ib}u_b=0$, where $T^{ab}$ is the usual Maxwell electromagnetic SET. 

We finally deduce that there is no radiation anywhere in the asymptotic region as the magnetic field is identically zero in the whole spacetime.

\subsection{Does a charge free-falling in a gravitational potential well radiate?}

A free-falling charge will radiate with respect to an observer at rest \cite{DeWittDeWitt1964,Boulware1980}, but this same charge will not radiate according to a comoving (free-falling) observer \cite{Bradbury1962}. In agreement with these results, an analysis based on our Rindlerian geometry above shows a net flux of energy in the asymptotic regions. This essentially involves transforming the Coulomb field of an inertial particle in Minkowski spacetime to Rindler coordinates. 
In the same manner this means that an accelerated observer will perceive a charge at rest in Minkowski spacetime as radiating. We can see here a classical analogue of the Unruh effect: an accelerated detector (e.g. an antenna) will detect radiation in the Coulomb field of a charge at rest (see e.g. a discussion along these lines in~\cite{PauriVallisneri1999}). 

In this paper we are concentrated in the simplest situation showing the subtleties of the presence of radiation: acceleration in rectilinear motion. However, let us just note here that a charged particle orbiting a planet in free-fall motion will also produce radiation as seen by static observers. So it is not the deviation of a trajectory from free fall what causes the presence of radiation.

\subsection{The nature of radiation}

The previous discussion leads to the following image. At least in the context of accelerating charges, to radiate or not to radiate is a perception issue. The Maxwell SET does not change its form from a radiating situation to a non-radiating one. In this sense, radiation is not encoded in an objective flow in the Maxwell SET. It is instead a matter of how one splits the SET into radiating and non-radiating parts, something that is beyond the SET itself. 

For instance, for inertial observers in Minkowski spacetime, and given an arbitrary trajectory for a point-like charge, one should follow Teitelboim and collaborators~\cite{Teitelboim1970,Teitelboimetal1980} and separate the electromagnetic field into 
a Coulomb part $F_{\rm C}^{ab}$ and a radiation part $F_{\rm R}^{ab}$ defined as
\begin{align}\label{Splitting}
F_{ab}&=F_{\rm C}^{ab}+ F_{\rm R}^{ab},
\nonumber\\ \nonumber\\ 
 F_{\rm C}^{ab}= \frac{2q}{(R^c u_c)^2} u^{(a}n^{b)}
\big|_{\rm ret},
&\qquad
F_{\rm R}^{ab}= \frac{-2q}{R^c u_c}
\left[(a^cn_c) u^{[a}n^{b]}+ a^{[a}n^{b]} \right]
\big|_{\rm ret}.
\end{align}
In this expressions we use the following notation: $R =(x-x_t)$ is the four-vector joining the spacetime point $x$ with the retarded position $x_t$ of the point-like charge; $-R^a u_a$, is the retarded distance to the charge; $n^a=-R^a/ (R^b u_b)$ is the retarded orientation; $(ab)$ and $[ab]$ indicate symmetrisation and anti-symmetrisation in the corresponding indices, respectively; and $|_{\rm ret}$ reminds that these expressions must be evaluated at the retarded time.    

While the radiation part of the field depends on the instantaneous retarded velocity and acceleration of the charge, the Coulombian part depends only on the instantaneous retarded velocity of the charge. From these quantities one can construct Maxwell's SET and split it into two parts
\begin{align}
T^{ab}=T_{\rm L}^{ab}+T_{\rm R}^{ab},
\end{align}
where the local $T_{\rm L}^{ab}$ and radiative $T_{\rm R}^{ab}$ are given by
\begin{align}
T_{\rm L}^{ab}:=T_{\rm CC}^{ab}+T_{\rm CR}^{ab}
,\qquad T_{\rm R}^{ab}:=T_{\rm RR}^{ab}.
\end{align}
The labels CC, CR, and RR represent the terms that come from products of the radiative and Coulombian parts of the electromagnetic field.

When a charge is accelerating $T_{\rm R}^{ab} \neq 0$ and it has been proved that it encodes all the radiative properties of the field. For instance, its $0i$ components are non-zero signalling a flux of energy travelling towards infinity~\cite{Teitelboim1970,Teitelboimetal1980}. The radiative part is conserved off the particle,
$\nabla_a T_{\rm R}^{ab}=0$, and has a Dirac's delta source at the particle itself. By looking only at this term one could interpret the radiation process as something that occurs locally, as an emission that starts from the particle itself, contrary to the previous Boulware explanation. There is no contradiction however. Teitelboim's splitting is explicitly of a retarded nature. In a region of uniform acceleration and close to the particle, one could have taken equivalently an advanced splitting leading to a different $T_{\rm R}^{ab}\neq0$ which now would contain just ingoing radiation. The nice feature of the retarded splitting when using retarded fields is that it is consistent with the emission of radiation for wave fronts arbitrarily far from the particle.

How does Teitelboim's splitting fit with the previously expressed idea that in an accelerating frame an accelerating charge does not radiate? The connection appears when one realises that an equivalent splitting can be performed using the acceleration as defined relative to the Rindlerian frame~\cite{Hirayama2001}. For instance, the field of a charge at a fixed distance in the Rindlerian spacetime will only have a Coulombian part 
\begin{align}
F^{ab}=F_{\widetilde{\rm C}}^{ab},
\end{align}
although, tensorially speaking, it is the same field as that of a uniformly accelerating charge in Minkowski spacetime, i.e.,
\begin{align}
F_{\widetilde{\rm C}}^{ab}=F_{\rm C}^{ab}+ F_{\rm R}^{ab}.
\end{align}
The Maxwell SET will just have the form
\begin{align}
T^{ab}=T_{\widetilde{\rm L}}^{ab}=T_{\widetilde{\rm C}\widetilde{\rm C}}^{ab},
\end{align}
having no radiation term. Tuned accelerated observers in Minkowski spacetime (by tuned we mean with Rindler accelerations, see below) will share this same perception of no radiation, with no splitting of the electromagnetic field. The Einstein (mechanical) equivalence principle is extended in this way to moving charges, finding no violations: by means of experiments with moving charges there is no way to tell whether a lab is accelerating or restrained from falling into a gravitational well.

It is also interesting to compare Dirac's and Teitelboim's definitions of the radiation field in a Minkowskian situation. In Teitelboim's splitting it appears as if the radiation were created at the particle itself. Instead, Dirac defined the radiation field as 
\begin{eqnarray}
F_{ab}^{\rm rad}:=F_{ab}^{\rm ret}-F_{ab}^{\rm adv}.
\end{eqnarray}
This definition makes the radiation field in the surroundings of a particle subject to uniform acceleration to be zero. Only when reaching  future null infinity the two definitions coincide. Dirac's definition conveys the idea that radiation only appears as a far field and cannot be distinguished close to the particle. Both definitions have nice features but none of them capture the actual relational nature of radiation.  

At this point let us make some further observations. If we were just considering classical electrodynamics with just the previous emission mechanism, we would not need to associate independent degrees of freedom to the electromagnetic field. One could always associate the presence of some radiation passing through a region as the result of some specific rearrangement of elementary particles somewhere else in combination with a relative perception mechanism. To use independent degrees of freedom for the electromagnetic field would just be a convenient way of working since in many applications one does not need to worry about the emission mechanism. Notice, however, that the situation changes when considering quantum mechanical effects. For example, the phenomenon of particle anti-particle annihilation can be taken as evidence that the electromagnetic field actually possesses independent degrees of freedom, with matter degrees of freedom being transmuted into electromagnetic ones.

In this paper we are only considering locally flat situations. The presence of spacetime curvature adds additional complications that we wanted to separate in order to have a clean discussion. 
The presence of back scattering (due to spacetime curvature) generates tails in the propagators which in turn hinder the naturalness of the splitting discussed above. For example, Villarroel~\cite{Villarroel1975} proposed a splitting in curved backgrounds but the radiation SET does not contain all the radiated energy. The situation when defining radiation in general relativistic settings is actually parallel with the relational notion of quantum particle in curved backgrounds~\cite{BirrellDavies1984, PauriVallisneri1999}.

\subsection{Radiation by composite particles}

Before ending this section let us also discuss the radiation emitted by a composite particle (or particle of finite size and internal structure). As we will see in the next section, this analysis is very relevant as one can only make physical sense of self-forces when going away from the point-particle assumption.  

Let us consider an extended charge-current field $J^a$ with total charge $q$ and whose spatial extension is of compact support. It can be interpreted as a charged object. The radiation generated by such an object could be extremely complicated, containing all sort of multipole components. It all depends on the internal complexity of the object. However, when thinking of a model for a system that effectively behaves as a point-like particle, we must assume that the composite system is as simple as possible. One would also like to be able to associate a single (sufficiently precise) effective trajectory to the composite particle. For these reasons, the most used models for a structured particle are a uniformly charged sphere and a spherical shell. The radiation produced by an extended object of this kind is approximately equal to that of an equivalent point charge only if we make an additional assumption: that the accelerations involved are very small as compared with the typical (inverse) size of the composite system, $g \ll 1/d$. 
In this scenario the fields originated at different locations of the composite object would not be able to interfere significantly at infinity, resulting in a radiation approximately equal to that that would have been produced by the charges separately. In summary, under the previous hypothesis, it is reasonable to expect that the radiation from a composite system can be very well approximated by that of a single point charge with the total charge of the composite and an effective average acceleration. We can say that the existence and quantity of radiation is robust in passing from the elementary to the simple composite system. Again, under the previous conditions, most of the radiation at infinity is concentrated around frequencies $\omega \sim g$. The condition $g \ll 1/d$ is telling us that the radiation is not coming from short scale characteristics of the composite object, but essentially from its motion as a whole.

On the contrary, as we will argue, self-forces are not equally robust. Self-forces explore higher frequency features of the fields and so, can in principle subtly depend on the structure of the composite. This can be seen even in the LAD equation~\eqref{Ren-LAD}, which depends on the $\dot{a}^b$ characteristics of the trajectory, while the radiation field depends only up to the $a^b$ features.

\subsection{Summary}
\label{Subsec:Summary}

As a synthesis of the history of this controversy we can say that all the main participants provided arguments with elements of truth. A uniformly accelerating particle in some sense radiates and in some other sense does not: the crucial ingredient is the relation between the trajectory of the particle and the global properties of the spacetime in which it evolves.

\section{The self-force equation}
\label{Sec:Selfforce}

The idea that the electromagnetic field produced by an accelerating charge should affect its own motion was realised by several researchers well before the special relativistic framework was developed~\cite{Poincare1894,Planck1896}. It is clear that one cannot directly deal with idealised point charges to analyse this back-reaction. The self-field diverges at the very position of the point-like particle making seemingly impossible to make any further assertion. Lorentz and Abraham realised that if a charge has a finite-size structure it is possible to envisage how do some self-force effects come about.

\subsection{Will a charge uniformly accelerating in Minkowski spacetime be subject to some self-force?}

A first intuitive analysis of a charge uniformly accelerating in Minkowski spacetime could lead us to believe that the presence of radiation at infinity would be accompanied by some local friction effect at the position of the charge itself. We will be using the word friction when thinking intuitively on a force that acts against the motion, i.e. proportional to the velocity and in the opposite direction. 
However, an equivalent intuitive analysis of a particle at rest in Rindler spacetime, with its corresponding absence of radiation at infinity, could make us believe that in the latter case the particle would not be subject to any friction force. As they stand, these two analysis are not compatible with one another. Our previous analysis of radiation based on a single elementary particle asserts that the situation is equal in both cases so the forces, if any, should also be equal in both cases. 

Indeed, the literature on the subject has apparently reached the consensus that the two situations are equal and that the intuitive analysis that turns out to be incorrect is that of the accelerated charge in Minkowski spacetime: in periods of constant acceleration the self-force vanishes and there are no friction forces at work. Indeed, the LAD self-force term~\eqref{Ren-LAD} vanishes for hyperbolic (constant proper acceleration $g$) trajectories. This can be easily seen in \eqref{Ren-LAD} noting that for a charge in hyperbolic motion both terms in the relativistic version are equal to $g^2 u^b$ and hence cancel out and is obvious in \eqref{Ren-LAD-nonrel}.

However, as mentioned in the introduction, this state of affairs leads to some interpretational problems. To understand the problem of the local energy budget, Fulton and Rohrlich~\cite{FultonRohrlich1960} elaborated on analyses by Schott~\cite{Schott1915} and proposed that the problem resides in an additional source of energy (and force) that typically passes unnoticed. This is an ``acceleration energy'' term, $Q=-2/3 q^2a^0$, which grows negative in regions of constant acceleration, thus compensating the energy extracted in the form of radiation. 
This term appears when writing the energy balance equation associated with the LAD equation~\eqref{Ren-LAD}, whose time component can be expressed as
\begin{align}
\frac{d(E+Q)}{d\tau } = -\frac{2}{3} q^2 g^2 u^0 + F^0_{\rm ext}.
\end{align}
Here $E$ is the dressed kinetic energy and $F^b_{\rm ext}$ is the external force that drives the charged body. The Schott acceleration energy $Q$ is reversible: it is accumulated during accelerated motion but returns to zero in inertial segments. Notice also that the acceleration energy does not show up as an addition to the inertial mass: it is neither in the radiation field nor inside the effective mass of the particle. This acceleration energy could be seen at first sight as mysterious and not very physical. 
However, in the 60s it was proved that it actually corresponds to the electromagnetic energy contained in the local term in the Maxwell SET~\cite{Coleman1961,Teitelboim1970}. 
In fact, in a very interesting paper~\cite{Rowe1978}, Rowe elaborated on previous works by Harish-Chandra~\cite{Harish-Chandra1945} and Weert~\cite{Weert1974} and proposed a new splitting of the Maxwell SET. This splitting is motivated by the different divergent properties of the terms and their distributional extension. More explicitly, attending to the different divergent properties of the different terms composing the local $T_{\rm L}^{ab}$ in Teitelboim's splitting~\eqref{Splitting}, it was separated into two terms different from the previous ones $T_{\rm CC}^{ab}$ and $T_{\rm CR}^{ab}$. In addition he provided a proper distributional definition of the expressions by adding appropriate delta contributions at the worldline of the point charge:
\begin{align}
  T_{\rm L}^{ab} &= T_{\rm sol}^{ab} + T_{\rm Schott}^{ab},\nonumber\\ \nonumber\\
T_{\rm sol}^{ab}&=\frac{q^2}{4\pi} \left[
\frac{1}{2} \eta^{ab} - \frac{2 u^{(a} R^{b)}}{R^d u_d}
-\frac{R^a R^b}{(R^d u_d)^2}+ \frac{2(R_c a^c) R^a R^b}{(R^d u_d)^2}
\right]\frac{1}{(R^e u_e)^4} 
-\frac14 q^2 \gamma^{-1} u^{b} a^{a} \delta^3(x-x_{\rm t}) ,
\nonumber\\
T_{\rm Schott}^{ab}&=\frac{q^2}{4\pi} \left[
2 a^{(a} R^{b)} -
\frac{2(R^c a_c) u^{(a} R^{b)}}{R^d u_d}
-4 \frac{(R^c a_c) R^a R^b}{(R^d u_d)^2}
\right]\frac{1}{(R^e u_e)^4}
-\frac23 q^2 \gamma^{-1} u^{a} a^{b} \delta^3(x-x_{\rm t}),
\label{SchottRowe}
\end{align}
where the label ``sol'' stands for solenoidal. Notice that the explicit delta-function terms appearing in the previous expressions are not symmetric in $ab$. In fact, they are there to eliminate other non-symmetric terms that appear when analysing the expression in a proper distributional manner \cite{Rowe1978}. Rowe's distributional definition of the three partial SETs, the radiative, the Schott, and the solenoidal parts, are indeed $ab$-symmetric both off-shell and on the worldline.  The specific way in which the delta functions are arranged has important consequences that we discuss below.

In our view, this splitting provides the cleanest interpretation we have seen in the literature. The first term is divergence free even at the particle position $\nabla_a T_{\rm sol}^{ab}=0$. The tensor $T_{\rm Schott}^{ab}$ is divergence free off the particle  while, at the particle position, provides a point source supplying precisely the instantaneously produced radiation that goes into the term $T_{\rm R}^{ab}$. This last term $T_{\rm R}^{ab}$ is conserved off-shell and has a source at the particle position of precisely the same form (but reversed sign) than the source in $\nabla_a T_{\rm Schott}^{ab}$. 
The tensor $T_{\rm Schott}^{ab}$ contains precisely the Schott (or acceleration) energy-momentum four-vector:  
\begin{align}
P^a= \int d\Sigma_b T_{\rm Schott}^{ab}= \frac23 q^2  {a}^a.
\end{align}
As a nice property let us mention that this integral does not depend on the hypersurface in which it is performed provided that it crosses the  trajectory of the particle at the same point. One could use for example any spacelike plane in Minkowski spacetime without worrying whether one or more particles intersect this plane orthogonally. Therefore, the acceleration energy-momentum is accumulated in a form of interference between the radiative and local field associated with an accelerating particle. 
In a series of papers~\cite{EriksenGron2000a,EriksenGron2000b,EriksenGron2000c,EriksenGron2002,EriksenGron2004} Eriksen and Gron   revised in detail the electrodynamics of a uniformly charged particle. In particular in~\cite{EriksenGron2002} they used  $T_{\rm Schott}^{ab}$ to analyse the localisation of the Schott energy-momentum. They showed that for a given time, the contribution to the Schott energy-momentum is zero from the region enclosed between any two concentric wave fronts which do not touch the position of a regularised extended particle. The Schott energy-momentum comes from a region surrounding the regularised particle which in the point-particle limit concentrates on the particle itself.

The previous splitting leads to the idea that at least part of the radiated energy (all of it in the regime of uniform acceleration) comes from a negative accumulation of acceleration energy. However, recalling that the splitting itself is observer dependent we are led instead to the idea that an equal energy budget is distributed in different but equivalent manners by different observers. An inertial observer in Minkowski spacetime will say that the emitted energy comes from an accumulation of acceleration energy. In a Rindlerian situation one would say instead that there is neither radiation nor acceleration energy, i.e. that only a Coulombian part will be present in the field.

\subsection{Will a charge free falling in a gravitational potential well be subject to some self-force?}

This situation was analysed by de Witt and Brehme~\cite{DeWittBrehme1960} for a particle free falling in a Schwarzschild geometry. They concluded that there exists some non-zero self-force effect but that in this case it is entirely due to the presence of tails in the propagator. This should not be present in a homogeneous gravitational field. Notice that our Rindlerian analysis avoids the presence of tails owing to the absence of curvature. 

Free falling in Rindler spacetime is equivalent to inertial motion in Minkowski. In this case it is reasonable to expect no self-forces. However, this might confront the fact that in this case there will be radiation at infinity. The acceleration energy notion comes to the rescue again. In this situation it is clear that Rindlerian observers have to assume that the negative acceleration energy $Q$ is being accumulated in a charge that is just moving inertially. This reinforces the idea that 
the acceleration energy, as well as all the other energies involved, depends on observational issues and do not have intrinsic local definitions.

\subsection{Non-uniform acceleration}
\label{Subsec:non-uniform}

From the LAD self-force \eqref{Ren-LAD} it results then that putting a charge in uniform acceleration only requires some extra work (associated with the radiated energy) in the transients. However, as mentioned in the introduction, the form of the LAD self-force in the transients is counter-intuitive. As we are going to explain, the cause of the interpretational problem comes from forgetting the inertial term in the LAD expression~({\ref{R-LAD}}). 

All derivations of the LAD equation involve an expansion in terms of derivatives of the acceleration, with the standard LAD expression maintaining only the first non-trivial term. The correct regime of application of the LAD self-force is then when $q^2 \dot{g} \ll m_{\rm ed} g$ (i.e. an adiabatic condition during the transients). Under this condition, when a charge accelerates we have   $-m_{\rm ed}a^i + \tilde{F}_{\rm S}^i \ll 0$. 

\begin{figure}
	\centering
	\includegraphics[width=.9\textwidth]{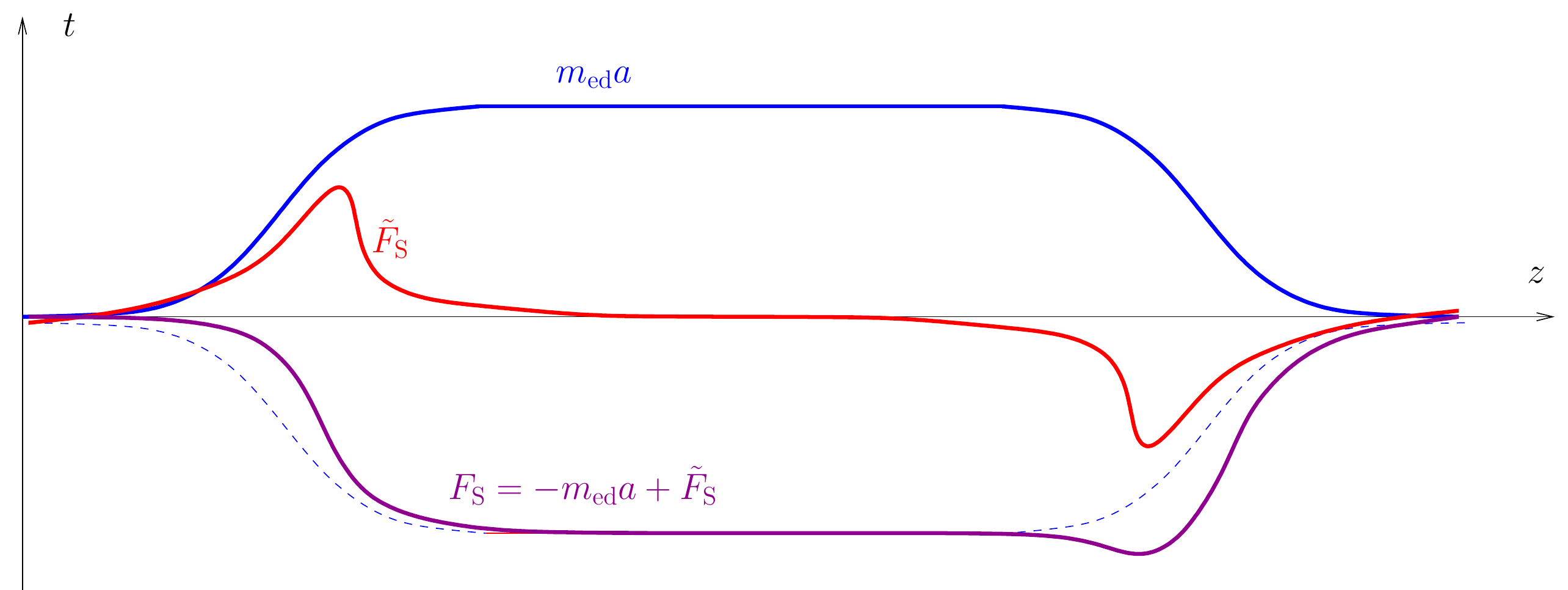}
	\caption{The diagram qualitatively portraits  the behaviour of the two terms $m_{\rm ed} a$ and $\tilde F_{\rm S}$ that form the full self-force $F_{\rm S}$ during a complete finite period of acceleration and deceleration $I_{\rm i}T_{\rm i}AT_{\rm f}I_{\rm f}$ depicted in figure \ref{Fig:Regimes}. }
	\label{Fig:Selfforce}
\end{figure}

Within the framework of an extended charge, what really happens in a transient $T_{\rm i}$ is that the total self-force acts against the acceleration. Furthermore, it increases until it stabilises at the value $-m_{\rm ed}a^i$ (see figure~\ref{Fig:Selfforce}). Therefore, it is not strictly correct to think that a charged extended particle has always an electrodynamic contribution added to the bare mass. This would correspond to a ``perturbative'' interpretation of the self-force (for instance, this is a potential interpretational problem of formal schemes like that on~\cite{Grallaetal2009}).

Here we maintain that a better ``non-perturbative'' interpretation is to consider that the electrodynamic inertial term appears progressively during the transient making it more difficult to accelerate the charged extended particle as compared with that with the charge off, so to speak. In the constant-acceleration regime $A$, the inertial term is all that remains and is the one responsible for making more difficult to accelerate a particle  when it is charged than when it is not: starting from two particles with equal bare masses, the one with charge acquires under acceleration an additional contribution to its mass. In practice this idea passes unnoticed because measurements of the inertial mass of a particle are performed in timescales larger than the typically very brief extended-particle crossing time, i.e. we measure the dressed mass.
Then, when comparing the behaviour of a charged particle with respect to an uncharged particle one takes two with equal dressed masses and conclude than they behave equally. 

When the extended charge starts recovering an inertial state (transient $T_{\rm f}$) the two terms of the self-force progressively disappear (figure~\ref{Fig:Selfforce}). The form of the LAD self-force indicates that the process during this transient $T_{\rm f}$ is not completely symmetrical with respect to that in $T_{\rm i}$  (see the change of sign in $\tilde{F}^i_{\rm S}$ in expression \eqref{Ren-LAD}). This asymmetry occurs because of the retarded nature of the self-force effect. During the transient $T_{\rm f}$ it is clear that the self-force is against the acceleration, that is, it helps recovering an inertial motion.

It is clear that the LAD self-force is not a frictional force in the sense of acting against the velocity of the particle. With hindsight it would have been difficult to understand that a self-force would have a frictional effect  proportional to the velocity but acting in opposite direction and also proportional to the square of the proper acceleration. A friction of this form would work against having a velocity and not against having an acceleration. While the former would have selected a preferred frame of reference, the latter is perfectly consistent with the idea that radiation reaction is just opposing non-inertial motion. As Lorentz himself appropriately put it, the self-force provides a {\em resistance} to acceleration~\cite{Lorentz1904}.

\section{A possible twist to the situation}
\label{Sec:Twist}

The image that results from the previous discussions is consistent and takes into account the knowledge on the topic accumulated during a century. However, one can still find at least two puzzling issues that suggest a interesting possible twist to the radiation-reaction problem.

\subsection{The rigidity hypothesis}
  
The first puzzling observation is related to the impossible rigidity of real extended bodies in relativity. The natural state of an extended body in Minkowski spacetime is inertial motion. In fact, when analysing physical situations that involve accelerations one typically imposes that the acceleration regime is preceded by a state of inertial motion. All the calculations we know of regarding extended charges explicitly or implicitly assume that the structure of the body is strictly rigid and that its charge is distributed with strict uniformity  assuming some shape (e.g. a rigid and uniformly charged sphere; see for instance~\cite{Rohrlich1999,Smorenburgetal2014}; in the latter the author reviews several rigid models comparing different approaches to the calculation of their behaviour). Rigidity is consistent with a regime of inertial motion and also with a regime of strict uniform proper acceleration throughout the body. Beyond that, rigidity does not make much sense or is restricted to very specific situations \cite{Born1909}. Moreover, it is well know that for non-uniform accelerations or in general relativity there is not even a well-defined notion of rigidity~\cite{MasonPooe1987}.

A transient regime $T_{\rm i}$ necessarily introduces tensions in a realistic extended body. But the problem permeates even when trying to produce uniform acceleration. On the one hand, a realistic extended albeit very small body will be constituted by a neutral atomic network uniformly sprinkled with charge excesses or deficits so that on average it results in a uniformly charged system. When applying an electric force to the system one is just pulling the charges which act as anchor points to pull the entire system. In any realistic situation in Minkowski spacetime one would be far from uniformly pulling the system. The structural forces within the system could keep it together but at the cost of continuous retarded readjustments of these forces. On the other hand, even if one considered that the uniformity of the charge is almost perfect, if one applies a constant force field to an extended charge the force tries to set each elementary charge into equal accelerations, not equal {\em proper} accelerations. But equal accelerations do not lead to a rigid acceleration. In a stable regime of uniform acceleration one needs that the distributions of accelerations through the extended charge is the very specific one consistent with a rigid object in relativity. A uniform force should produce instead a disrupting stretching of the structure which the internal structural forces (whatever their nature) would try to counteract.

Let us better illustrate  the previous discussion with the simplest extended system one can think of: two particles of charge $q/2$ and mass $m/2$ separated by a small distance and tied together by a spring of some sort (in~\cite{Lyle2010} the reader can find a compelling set of calculations involving this simple situation). Imagine that they are initially at rest and located at a distance $d_{\rm i}$ from each other. For the two charges to remain in this initial stable configuration, they have to be tied together in some way so that the electric repulsion is counteracted. That is why we put a spring connecting them. Now, let us accelerate the two charges in the direction in which they are connected (e.g.  the $z$-axis). If they accelerate equally, we know that the proper distance as seen from a reference frame instantaneously at rest with the charges is now $d>d_{\rm i}$. Therefore, if the rope connecting them were not elastic, it would break (this constitutes the so-called Bell's paradox~\cite{Dewan1959,Bell1976}). 

On the other hand, if we set up the two charges to follow precisely uniform acceleration trajectories satisfying 
\begin{equation}
    d_{\rm i}=\frac{1}{g_{\rm h}}-\frac{1}{g_{\rm t}},
\end{equation}
with $g_{\rm h},g_{\rm t}$ the accelerations of the head and tail charges, respectively, then the distance between them as seen from each charge is kept constant. In this case, and only in this case, the forces maintaining the charges together are just those present  in the initial configuration; the motion does not affect these forces.
These specific trajectories precisely correspond to the different rest positions in Rindler space. 

For these trajectories it is interesting to calculate the electromagnetic forces exerted by each charge on the other. For that, one has to use the form of the electromagnetic field produced by one point charge on the position of the other:
\begin{equation}
    F_{ab}=\frac{\mu_0 c}{4\pi}\left[\frac{q/2}{R^c u_c}\frac{d}{d\tau}\left(\frac{R_a u_b -R_b u_a}{R^c u_c}\right)\right]_{\rm ret}~,
\end{equation}
where $R^a$ denotes the retarded distance between both charges. The force is then calculated as $f_a=F_{ab}j^b$, with $j^b=(q/2)\gamma^{-1} u^b$ being the charge current of the charge that suffers the force. It is easy to obtain the two reciprocal forces between the particles. Defining an average acceleration as  
\begin{equation}
\frac{2}{g}=\frac{1}{g_{\rm h}}+\frac{1}{g_{\rm t}}
\end{equation}
we obtain the forces exerted by the tail charge on the head one and vice versa:
\begin{equation}
f_z^{{\rm t} \rightarrow {\rm h}}=\frac{\mu_0 c q^2}{16\pi d^2}
\left(1-\frac{dg}{2}\right)^2 
, \qquad 
f_z^{{\rm t} \leftarrow {\rm h}}=-\frac{\mu_0 c q^2}{16\pi d^2}
\left(1+\frac{dg}{2}\right)^2.
\end{equation} 
These relativistic forces have the structure of constant proper forces. One the one hand, we see that $|f_z^{{\rm t} \rightarrow {\rm h}}|<|f_z^{{\rm t} \leftarrow {\rm h}}|$, which means that the acceleration of the system causes a force opposing the acceleration itself. On the other hand, we see that the addition of the two forces (as if applied to the central point) results in a total force
\begin{equation}
f_z=-m_{\rm ed}g
,\qquad 
m_{\rm ed} = \frac{\mu_0 c q^2}{8\pi d} + {\cal O}(d^0).
\end{equation}
This term, that appears only in this uniformly proper acceleration regime, can be absorbed into the definition of an inertial mass, but as explained in section~\ref{Subsec:non-uniform}, its appearance is the very radiation reaction effect we should not forget. The mass
$m_{\rm ed}$ correspond to an electrodynamic energy which diverges in the $d \to 0$ limit. For real extended systems, it always stays finite.

In the generic case in which the accelerations did not follow this precise uniform pattern, the distances from one particle to the other as seen by each particle do not even coincide! They do not share an instantaneous reference frame. In both cases of non-uniform accelerations and of equal accelerations for head and tail charges, the structural forces maintaining the charges bound together will experience adjustments.

This toy system also illustrates an additional issue about which we have not said anything yet: the composition and behaviour of the spring (in fact it is difficult to say much about it; see next section). For example, as a material system it should also have mass. Then, the forces applied to the charges would be pulling from the spring making it to react in specific ways. For sure its reactions would not constitute a strict rigid motion.

\begin{figure}
	\centering
	\includegraphics[height=.5\textwidth]{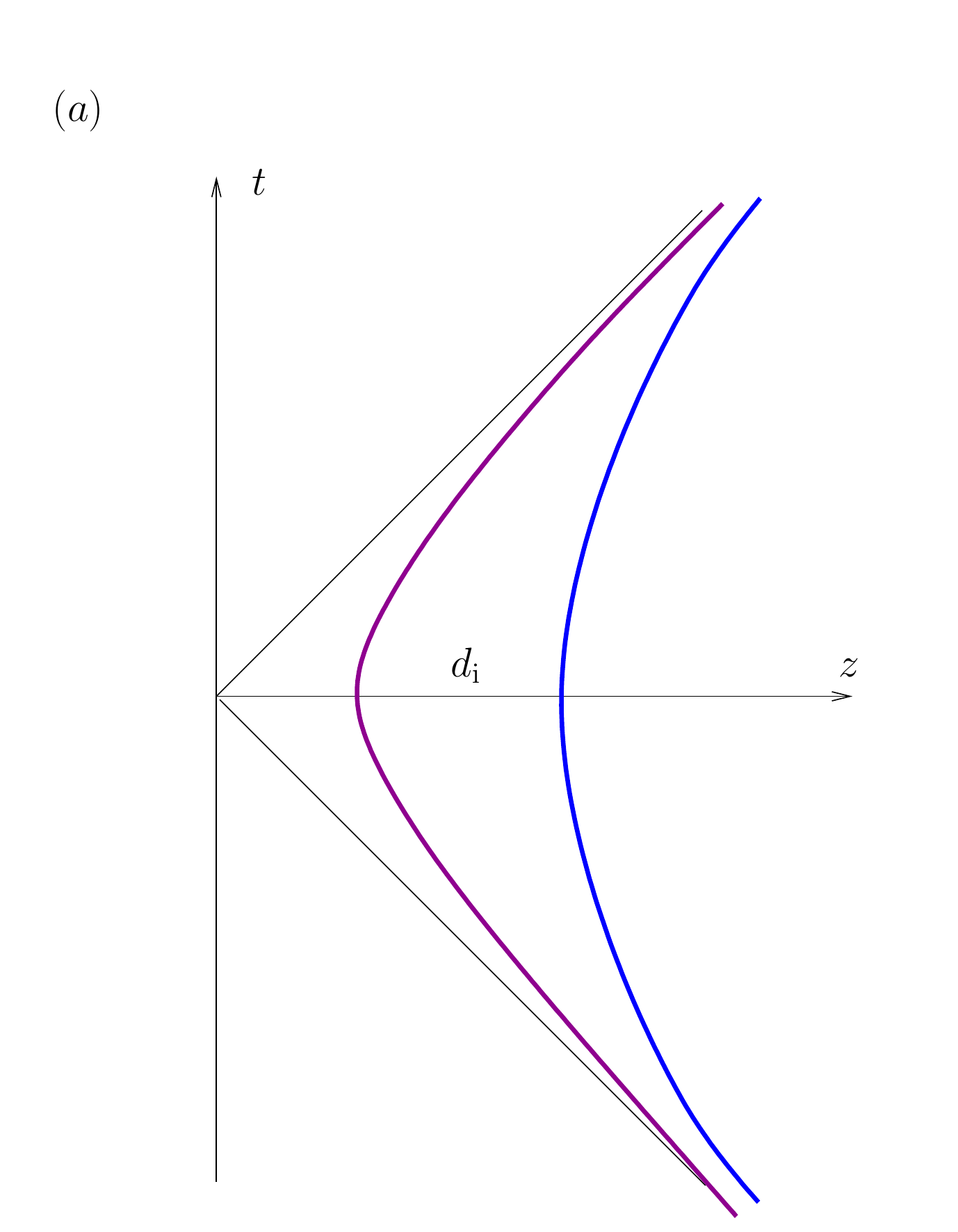}
	\quad	\includegraphics[height=.5\textwidth]{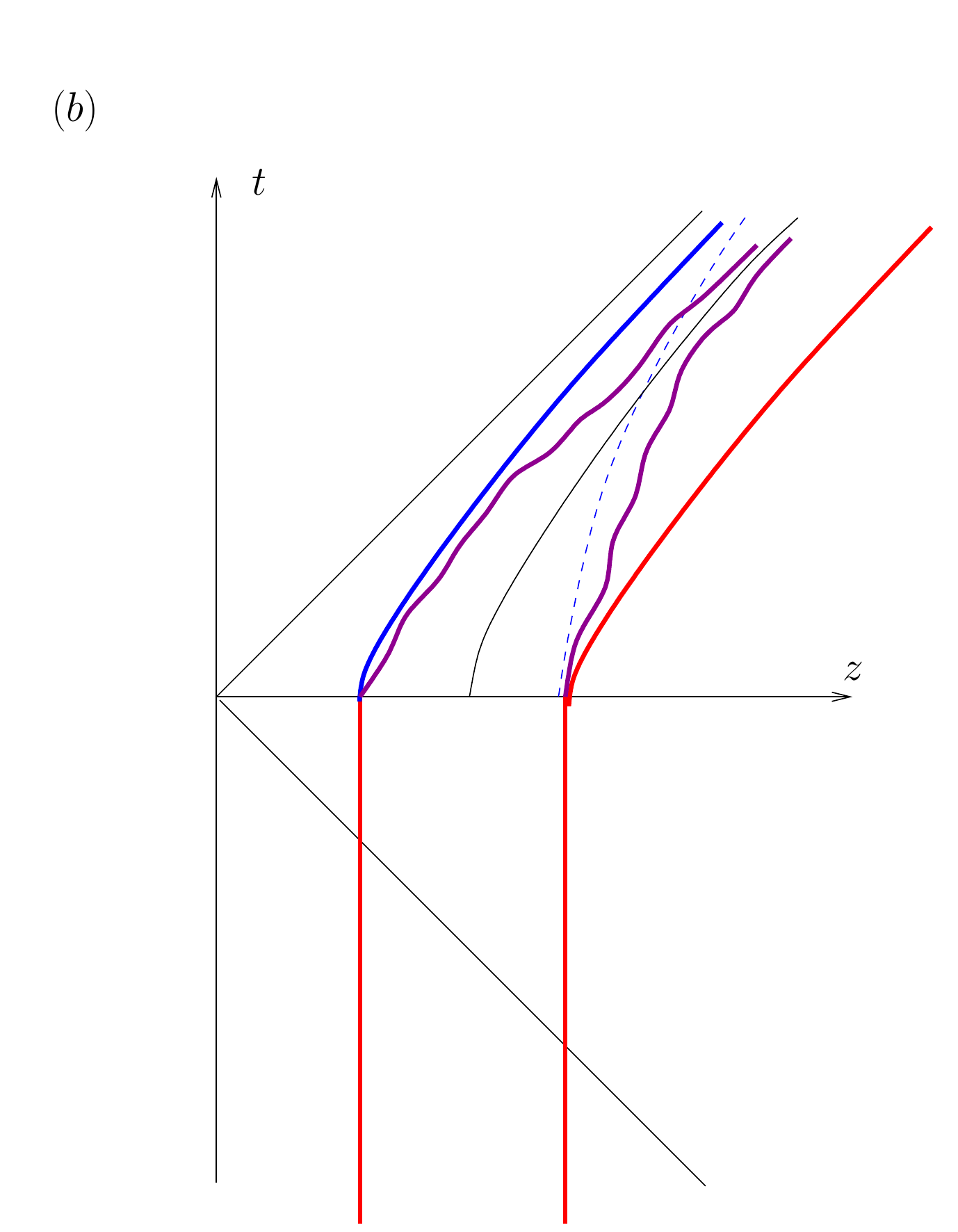}
	\caption{\emph{(a)} Two hyperbolic trajectories in Minkowski spacetime, or equivalently, two particles at rest in Rindler spacetime at different positions. The proper distance between these two trajectories is well defined and fixed to a value $d_{\rm i}$. \emph{(b)} A constant force field will try to set the two charges to follow equal hyperbolas, not the two unequal Rindlerian hyperbolas of figure~(a). The presence of a material spring joining the two charges may result in an oscillating trajectory instead of the rigid trajectory of figure~(a).}
	\label{Fig:MinkRindRigid}
\end{figure}

So, in setting an acceleration regime starting from an initial inertial motion there will always be a tension between the disrupting effect of the external forces and the structural forces that try to keep rigidity. 
For all the reasons explained, it is difficult to hold that by applying a constant force the toy-model extended system will move like in figure~\ref{Fig:MinkRindRigid}a, in which both charges undergo hyperbolic motion with different accelerations. It is more sensible to expect that qualitatively the system will move more like in figure~\ref{Fig:MinkRindRigid}b, where oscillations are present throughout the trajectory.

%

\subsection{Schott's energy and tensions}

The other puzzle comes about when rethinking the Schott term in the self-force \eqref{R-LAD}. Working in the point-like limit the acceleration energy grows continuously and exponentially in periods of uniform acceleration. The same happens with the Schott force term $da^i/d\tau$. As explained before, the cross term in Maxwell's SET contains the Schott energy-momentum. In the same manner it contains some pressure terms.
As shown in~\cite{EriksenGron2002} the Schott acceleration energy-momentum is localised essentially at the position of the particle itself. Looking at the Schott SET~(\ref{SchottRowe}) one can also check that, in a long period of uniform acceleration, large acceleration pressures accumulate at the location of the particle and its surrounding regions. These pressures should be compensated by the structural forces of the charged body in order to maintain its structural stability.

It is interesting to realise that one way in which one would be able to tame the accumulation of large acceleration energies and pressures is by having oscillations in the acceleration. Specifically, instead of accelerating uniformly, imagine that the system is effectively experiencing intermittent periods of acceleration essentially composed of a sequence of transients of the form $T_{{\rm f}j}T_{{\rm i}j}$, $j=1\cdots N$, so that the total process would read 
\begin{eqnarray}
I_{\rm i} T_{\rm i} T_{{\rm f}1} T_{{\rm i}1}T_{{\rm f}2} \cdots T_{{\rm f}N} T_{{\rm i}N} T_{\rm f} I_{\rm f}.
\end{eqnarray}
There are no periods of strict uniform acceleration. The notion of uniform acceleration appears only on average. Applying this idea to the system of two charges described before, one could image a situation similar to the one illustrated qualitatively in figure~\ref{Fig:MinkRindRigid}b. In subsection~\ref{Subsec:Model} below we will reinforce this possibility by working out a classical model of two masses bound together by a spring.

\subsection{An alternative view on radiation-reaction}

The previous two observations lead to an interesting conclusion. A pure rigid acceleration trajectory for an extended body is clearly physically unreasonable as we have discussed. On the other hand, an oscillating version (typical of elastic bodies), which on average might appear as indistinguishable from the former, can produce a much more intuitive interpretation of the emission and radiation-reaction effects. In this alternative view, in a period of uniform acceleration (on average) the system is emitting continuously (on average) and is back-reacted by a radiation-reaction force also continuously (again, on average). This conceptualisation could also avoid large accumulations of both accelerating energies and pressures. If the individual charges composing the extended system periodically went through periods of zero acceleration, then the Schott term  would oscillate, passing many times through zero without entering exponentially increasing regimes.

Having this picture in mind, it is also interesting to notice that internal oscillations of an extended body might, at least qualitatively, be simulated by a single trajectory of a point-like charge with added microscopic oscillations. In this way the standard LAD self-force expression could be formally used without encountering the interpretational puzzles associated with strictly uniform accelerations.
 
In fact, as we mentioned in the introduction, the interpretational problem with uniform accelerations has a parallel in the Unruh effect: does the coupling of a uniformly accelerating detector to a field cause the emission of field quanta? In periods of uniform acceleration it appears that there is no emission of particles~\cite{Grove1986,Crispinoetal2008}. In trying to understand this puzzle in more detail Parentani~\cite{Parentani1996} analysed a model system in which the trajectory of the detector was also treated quantum mechanically. He concluded that the periods of uniform acceleration actually have a micro-oscillating structure. In this case, emission of quantum photons involve recoil effects that perturb the trajectory. Our proposal here could be taken as a classical analogue of that model in electrodynamics.

\subsection{On the difficulty of producing a model of oscillations}
\label{Subsec:Model}

So far we have discussed two puzzling observations occurring in the standard treatment and have suggested a possible alternative way in which extended systems might turn out to behave when applying a constant force. But, can we prove that this alternative version is actually correct? Can we at least provide an exact model calculation corroborating this behaviour?
Here is when we face various difficulties. Let us mention a few without claiming to be exhaustive.
\begin{itemize}
\item
{\em Difficulty of ascribing a centre of mass/energy to a composed or extended system:}
Without a rigidity hypothesis it is not straightforward to ascribe a single trajectory even to the simplest composed system consisting  of just two particles~\cite{Pryce1948}.
 
\item
{\em Difficulty of introducing interactions between relativistic point-particles:}
When trying to construct a simple model for a composed system one could think of two charges bound together by a spring (or interaction) of some sort. However constructing a model for interactions between relativistic particles encounters important obstacles~\cite{Currieetal1963}.

\item
{\em Difficulty of treating bounded systems in electrodynamics:}
Modern physics is built upon the idea that a consistent relativistic treatment of a system of elementary particles and electromagnetic fields requires to treat them all as quantum fields. But the problem then is that although the theory seems well defined, to calculate even the simplest situation (other than just scattering amplitudes) needs approximations of different sorts. For instance, the complex situations one encounters in condensed matter systems are typically confronted (in many cases with great success) by using non-relativistic quantum mechanics.    

\end{itemize}
As a summary, it appears that the framework that should allow for a consistent treatment of composed relativistic systems is still too difficult to control; and, on the other hand, the simple models one tries to build to effectively describe the more complex situations have important conceptual problems to deal with.

At this stage we do not know how to solve these difficulties to produce neither a realistic nor a simple relativistic model of the situation (but this does not mean that we should take the mathematically controllable situation as the one providing the physically correct picture). What we can do here is to work out a simple analogue classical model that exhibits oscillations of the form we suggest might exist.

A small but macroscopic system with total charge $q$ could be composed of zillions of charged particles (electrons, protons). The total charge is provided by a small mismatch between the number of protons and electrons in the structure: typically a surplus or a deficit of electrons in an otherwise neutral atomic network. As a classical image of the system we can image it as having a uniform distribution of mass sprinkled with points of charge. When applying a constant force field to the system, these points of charge will act as anchor points  which can be used to  pull the entire system. Let us take this image to the bones and consider a system of two particles with mass $m/2$ located at $x_{\rm t}$ and $x_{\rm h}>x_{\rm t}$. The masses are connected by a spring of natural length $b$ and constant $k$. The head particle at $x_{\rm h}$ has a charge $q$ to which we can apply a constant electric force from an initial time $t=0$ on. The tail particle however does not have a charge. The equations of motion of this system can be written in the form:   
\begin{align}
&\frac{m}{2} \ddot{x}_{\rm h}=f-k(x_{\rm h}-x_{\rm t}-b),
\qquad
\frac{m}{2} \ddot{x}_{\rm t}=k(x_{\rm h}-x_{\rm t}-b) .
\end{align}
These equations can be easily be solved for the initial conditions at $t=0$ that the two particles are at rest at positions $x_h(0)=b/2$, $x_t(0)=-b/2$  leading to
\begin{align}
x_{\rm h}&=\frac{b}{2}+ \frac{f}{2m} t^2
+ \frac{f}{4k } \left[1-\cos(2\sqrt{k/m}\,t)\right],
\nonumber\\
x_{\rm t}&=-\frac{b}{2} + \frac{f}{2m} t^2
- \frac{f}{4k } \left[1-\cos(2\sqrt{k/m}\,t)\right].
\end{align}
The accelerations of the two masses are respectively
\begin{align}
 \ddot{x}_{\rm h}= (f/m)\left[1+\cos(2\sqrt{k/m}\,t) \right],\qquad
 \ddot{x}_{\rm t}= (f/m)\left[1-\cos(2\sqrt{k/m}\,t) \right].
\end{align}
We clearly see that the head particle starts accelerating with acceleration $\ddot{x}_{\rm h}(0)=2f/m$, that is, as if it was not connected to anything else. Progressively this acceleration diminishes owing to the pull of the tail particle and enters an oscillatory regime passing by periodic moments of zero acceleration. On the other hand, the centre of mass of the system accelerates uniformly with acceleration $[\ddot{x}_{\rm h}(0)+\ddot{x}_{\rm t}(0]/2=f/m$.

\subsection{Back to the equivalence principle}

But what happens now with the analogous situation of an extended charge at rest in a Rindlerian spacetime? First of all, we have to realise that the acceleration structure of Rindler spacetime is such that it naturally produces different proper accelerations at different distances from the domain wall. For an extended structure at rest the acceleration structure is precisely the one that leads to pure rigidity. Rindler forces do not naturally lead to oscillations within the extended charge. Of course, the structural pressures have to maintain the static form of the extended charge. The difference with the previous situation is that now naturalness does not impose that initial conditions should be inertial motion. The presence of a domain wall in this spacetime makes conditions in which the distance to the wall are kept fixed perfectly reasonable.

The image that results from this discussion is that Minkowski spacetime and a domain wall spacetime may translate its global properties into different natural internal structures for the extended particles living in them. In this way we have that the electrodynamics formalism itself does preserve the equivalence principle but this might be broken by the different natural initial states on both situations. The situation can be seen as analogous to that  of general relativity in cosmology: although relativity builds upon the idea that one cannot distinguish between different inertial states, in practice the presence of the cosmic microwave background introduces a natural rest frame with specific effects.

\section{Summary and conclusions}

The classical electromagnetic radiation-reaction problem has attracted
the attention of many researchers for  more than one hundred
years\footnote{50 years ago Ginzburg already coined this a ``perpetual
problem''~\cite{Ginzburg1970} and thought to settle the issue.}. It is
the first instance of the potential clash between having point-like
discrete objects coexisting in interaction with continuous fields. Many
notions of modern quantum field theory have their roots in this
apparently simple problem.

In this work, first we have revised the literature on the classical
electromagnetic radiation-reaction problem seeking to understand several
questions that may appear paradoxical in a first look at the problem:
\begin{itemize}
    \item Does a uniformly accelerated charge in Minkowski spacetime radiate? 
\item Does an equivalent charge maintained in a fixed position
on a gravitational well radiate? 
\item Is the self-force a friction force or a proper
acceleration resistance force? 
\item Does the self-force produce some
backreaction on a particle in regimes of uniform acceleration?

\end{itemize}

Our revision has been useful to fully appreciate that the emission of
radiation is an observer dependent issue. It complements other
discussions one can find in the literature (see
e.g.~\cite{PauriVallisneri1999}). The emission of radiation is not
encoded in any stress-energy tensor but in the way one inquires into it.
In this sense, it appears parallel to the blurred notion of particle in
curved spacetimes~(see e.g. \cite{BirrellDavies1984}). At least in the classical theory,
radiation within a system of charged particles is an exclusively
relational notion. We have also shown how this notion of radiation fits the existence of self-forces. As a synthesis, different observational perspectives will make people 
analyse the energy budget in different ways. For example, an inertial
observer will say that a uniformly accelerated charge generates some
radiating energy plus some acceleration (or Schott) energy. In turn, an
accelerating observer will say that the only energy present is a
Coulombian contribution to the inertial mass of the charge. Our
presentation also advocates a separation between any bare mass the
charged system may have and an electromagnetic contribution. In this way
it is easy to appreciate that the self-force is a force resisting
acceleration, that is, changes from inertial motion.

We hope our presentation up to section~\ref{Sec:Selfforce} will help
improving the access to the relevant information by new generations of
curious people. In section~\ref{Sec:Twist}
however we take an step further and propose a rethinking of the previous
standard paradigm. We put forward the idea that precisely the relational
connection between the motion of a charged body (small but with
an actual internal structure) and the global characteristics of the
spacetime it inhabits, should tend to excite, in some circumstances,
internal vibrational degrees of freedom of the body. This would happen
whenever an external electromagnetic force is used to modify a natural state of motion
in the background spacetime. Note that this ``natural'' state of motion need not be
geodesic motion but is more related with a notion of acceleration with respect to the global features of the spacetime (i.e. asymptotic regions and matter content alike). For instance, we argue
that a constant external force acting on a charge body in Minkowski
spacetime could make it internally oscillate inhibiting the
generation of exponentially large Schott (acceleration) energies. While
the total emitted radiation will be equal to that of a structureless
uniformly accelerated charge, in this case the radiation will contain
periodic fluctuations. We then argue that the situation would be
different for the same body fixed in a gravitational well. Then, it is
more reasonable to expect that the vibrational degrees of freedom will
remain unexcited. Thus, this phenomenon might allow to differentiate the
two situations. As a result it is as if we were effectively  breaking the equivalence principle. The global characteristics of
the spacetime would have imprinted some natural initial conditions on
the internal states of the bodies.

We have highlighted the difficulties in producing a solvable model
exhibiting the described behaviour. We only have been able to collect
arguments in favour of this alternative paradigm, including the
formulation of a very simple solvable classical model that exhibits
precisely the advocated characteristics. We feel that to know about this
open possibility is interesting and could open new research trails.

\acknowledgments
Financial support was provided by the Spanish Government through the projects FIS2017-86497-C2-1-P, FIS2017-86497-C2-2-P (with FEDER contribution),
FIS2016-78859-P (AEI/FEDER,UE), and by the Junta de Andaluc\'{\i}a through the project FQM219. CB acknowledges financial support from the State Agency for Research of the
Spanish MCIU through the ``Center of Excellence Severo Ochoa'' award to the Instituto de Astrof\'{\i}sica de Andaluc\'{\i}a (SEV-2017-0709).

\bibliography{biblio-on-the-radiation-reaction-problem}

\end{document}